\documentclass[%
 aip,
 amsmath,amssymb,
 reprint,%
]{revtex4-1}

\usepackage{graphicx}
\usepackage{dcolumn}
\usepackage{bm}

\usepackage[utf8]{inputenc}
\usepackage[T1]{fontenc}
\usepackage{mathptmx}

\usepackage[version=3]{mhchem} 

\begin{document}


\title{Single photon emission and single spin coherence \\of a nitrogen vacancy centre encapsulated in silicon nitride}

\author{Joe Smith}
\email{j.smith@bristol.ac.uk}
\thanks{Contributed equally to this work.}
\affiliation{QET Labs, Department of Electrical and Electronic Engineering and H. H. Wills Physics Laboratory, University of Bristol, Bristol BS8 1UB, UK}
\affiliation
{Quantum Engineering Centre for Doctoral Training, Centre for Nanoscience \& Quantum Information, University of Bristol, Bristol BS8 1FD, UK}

\author{ Jorge Monroy-Ruz}
\thanks{Contributed equally to this work.}
\affiliation{QET Labs, Department of Electrical and Electronic Engineering and H. H. Wills Physics Laboratory, University of Bristol, Bristol BS8 1UB, UK}
\affiliation
{Quantum Engineering Centre for Doctoral Training, Centre for Nanoscience \& Quantum Information, University of Bristol, Bristol BS8 1FD, UK}

\author{John G. Rarity}
\affiliation{QET Labs, Department of Electrical and Electronic Engineering and H. H. Wills Physics Laboratory, University of Bristol, Bristol BS8 1UB, UK}

\author{Krishna C. Balram}
\affiliation{QET Labs, Department of Electrical and Electronic Engineering and H. H. Wills Physics Laboratory, University of Bristol, Bristol BS8 1UB, UK}

\date{\today}

\begin{abstract}
Finding the right material platform for engineering efficient photonic interfaces to solid state emitters has been a long-standing bottleneck for scaling up solid state quantum systems. In this work, we demonstrate that nitrogen rich silicon nitride, with its low auto-fluorescence at visible wavelengths, is a viable quantum photonics platform by showing that nitrogen vacancy centres embedded in nanodiamonds preserve both their quantum optical and spin properties post-encapsulation. Given the variety of high-performance photonic components already demonstrated in silicon nitride, our work opens up a promising avenue for building integrated photonic platforms using solid state emitters.  

\end{abstract}

\maketitle
The ability to isolate and manipulate single atom-like systems in the solid state has heralded a new age of quantum engineering. Given the increased decoherence that these atom-like systems exhibit in comparison to real atoms, the primary motivation for sustained research interest arises mainly from the prospect of building scalable platforms, with efficient photonic interfaces using nanofabrication. As the semiconductor industry has repeatedly shown, integrated chip-scale platforms can vastly exceed the complexity and connectivity of discrete devices. Building on the progress in silicon photonics, manipulation of quantum states of light in chip-scale platforms have also grown in their scale and complexity in the last decade \cite{wang2018multidimensional}. Adding solid state atom-like systems to these integrated quantum photonic platforms will help us realise deterministic single photon generators \cite{lindner2009proposal} and on-chip quantum memories \cite{childress2013diamond}, two crucial components for photonic quantum technologies.

A variety of candidate atom-like platforms have been pursued for integration with on-chip waveguides. These range from InAs quantum dots in GaAs waveguides \cite{lodahl2015interfacing}, colour centres in diamond and silicon carbide \cite{awschalom2018quantum}, 2D materials \cite{xia2014two}, to dye molecules \cite{Polisseni:16} and rare earth ions in glass \cite{zhong2015nanophotonic}. Amongst these platforms the nitrogen vacancy (NV) centre in diamond is unique in possessing a long-lived ($\sim {\mu}s$) spin coherence that can be optically manipulated and read-out at room temperature \cite{knowles2014observing}. Demonstrations of nuclear spin entanglement to error correct the NV centre spin \cite{hirose2016coherent} and entanglement between distant NV centres \cite{bernien2013heralded} show the possibility of engineering complex quantum information processing platforms around the NV centre. These pioneering experiments have relied on monolithic diamond samples with photonic interfaces provided by bulk optics, with limited scope for scalability. To realise the full potential of the NV centre in quantum information processing, it is apparent that integrating NV centres with on-chip photonic waveguides and cavities is essential. 

Diamond, which serves as the host for NV centres, is notoriously difficult to etch on account of its inert, rigid carbon lattice. While there have been demonstrations of nanophotonic devices in bulk diamond \cite{faraon2012coupling,sipahigil2016integrated}, using reactive ion etching in an oxygen plasma, it is difficult to foresee building large scale quantum circuits using this platform. An alternative approach relies on using NV centres located in nanodiamonds (size 10-20 nm) formed by milling bulk diamond. By embedding these nanodiamonds in high-index (relative to fused silica) dielectric films, it would be possible to realise a photonic waveguiding platform. The dielectric material would need to satisfy the following criteria for quantum information: 1) low absorption in the visible wavelength range, 2) low auto-fluorescence to preserve the NV single photon statistics, 3) preserve the intrinsic radiative quantum efficiency, 4) minimise added dephasing effects on the NV electron spin and 5) use established fabrication procedures for building integrated circuits. Criteria 1-4 are essential for building a quantum photonics platform, while 5 is key for large scale quantum integration (LSQI).

The three most viable dielectric thin film platforms for building quantum photonic circuits encapsulating NV centres hosted in nanodiamonds are: titania (\ce{TiO_2}) \cite{choy2012integrated}, aluminium nitride (\ce{AlN}) \cite{pernice2012high} and silicon nitride (\ce{Si3N4}). Amongst these, \ce{Si3N4} is the most attractive, given the impressive performance already achieved in integrated photonics, such as broadband frequency combs \cite{kippenberg2011microresonator} and integrated frequency converters\cite{li2016efficient}, and being mature in terms of foundry compatibility \cite{domenech2018pix4life,stroganovsilicon}. Unfortunately, stoichiometric silicon nitride  (\ce{Si3N4}) has broad auto-fluorescence around the zero phonon line of the NV centre (637 nm), which adds background noise and makes it challenging to observe the quantum signature (antibunching) of the NV centre single photon emission \cite{mouradian2015scalable}. Therefore, for silicon nitride to be a viable quantum photonics platform compatible with NV centres, this background auto-fluorescence needs to be suppressed. In this paper we build on the idea of using non-stoichiometric films \cite{Cernansky2018,Gorin2008} to minimise this background auto-fluorescence.  In contrast to evanescent coupling approaches, here we demonstrate that single emitters can be encapsulated in a high-index dielectric medium (n $>$ 1.9), which allows us to engineer greater overlap and achieve stronger interaction between the dipole emitter and the waveguide mode. To the best of our knowledge, this technique has been demonstrated only with low index contrast platforms, which are not suitable for nanophotonics \cite{schell2013three,khalid2015lifetime,hui2019all}. We show that nitrogen-rich amorphous silicon nitride films ($\alpha$\ce{SiNx}) can serve as a viable photonic platform for interfacing with NV centres located in nanodiamonds by demonstrating single photon emission and single spin coherence in nanodiamonds that are capped with 100 nm of $\alpha$\ce{SiNx}. In contrast to probing the film for encapsulated emitters, here, we characterise the optical and spin properties of the same single NV centre(s) (identified using fiduciary markers) before and after film deposition allowing us to unambiguously demonstrate that the NV centre survives the plasma deposition process and that nitrogen-rich $\alpha$\ce{SiNx} provides a quantum photonics platform compatible with single atom-like systems.

\begin{figure}
 \includegraphics[width=0.45\textwidth]{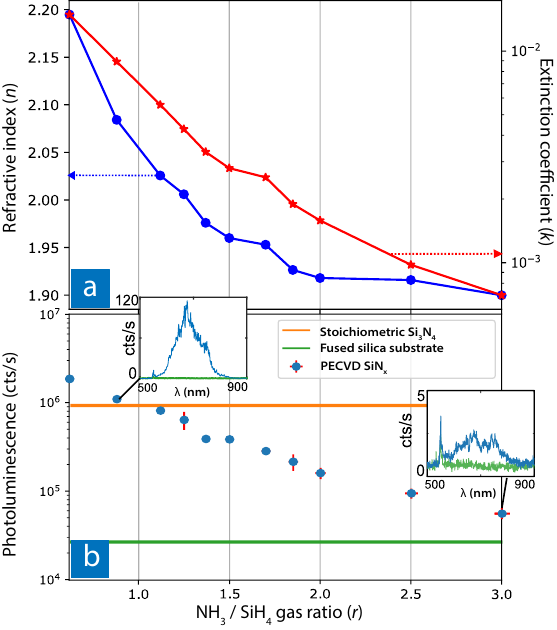}
  \caption{a) Refractive index (n) and absorption coefficient (k) of $\alpha$\ce{SiNx} films (@ $\lambda = 637$ nm) with increasing ammonia (\ce{NH3}) to silane (\ce{SiH4}) gas ratios ($r$) deposited using plasma enhanced chemical vapour deposition (PECVD). Values from Cauchy model fit to ellipsometry data for the transparent region ($\lambda > 500$ nm)   b) Integrated photoluminescence (PL) spectrum for the samples in (a) as a function of gas ratio. Orange and green lines represent the PL of a stoichiometric \ce{Si3N4} film deposited using low pressure chemical vapour deposition (LPCVD) and background PL measured on a bare fused silica substrate, respectively. Insets show representative PL spectra (blue) for $\alpha$\ce{SiNx} samples with low (left) and high nitrogen content (right). Here, the background PL from bare fused silica is shown in green for reference. Vertical error bars obtained by averaging integrated values for repeat measurements. Horizontal error bars correspond to uncertainty in the gas flow controller.}
  \label{fgr:nitride}
\end{figure}


Increasing the nitrogen content of $\alpha$\ce{SiNx} films helps reduce the background photoluminescence (PL) emission \cite{Cernansky2018,Gorin2008}. For technological viability, the key question to address is whether the background PL can be reduced sufficiently to observe single photon emission statistics from an NV centre, encapsulated with $\alpha$\ce{SiNx}. In addition, given the prevalence of N atoms in $\alpha$\ce{SiNx}, one needs to measure the spin properties of the NV centre to ensure that the electron spin coherence is not significantly reduced after film encapsulation. To study the effect of $\alpha$\ce{SiNx} on encapsulated NV centres, we start by characterising the background PL emission from $\alpha$\ce{SiNx} with varying nitrogen content. To achieve this, 300 nm amorphous silicon nitride films ($\alpha$\ce{SiNx}) were deposited on fused silica substrates using plasma enhanced chemical vapour deposition (PECVD). The $\alpha$\ce{SiNx} films were deposited varying the \ce{NH3} to \ce{SiH4} gas flow ratio $R = [\ce{NH3}]/[\ce{SiH4}]$ from 0.6 to 3.0 while keeping the total gas flow constant as well as the chamber pressure at 1.0 Torr and the substrate temperature fixed at 300$^\circ$ C.

After deposition, the refractive index ($n+ik$) of the films was extracted using ellipsometry, as plotted in Fig \ref{fgr:nitride} (a). As shown in Fig \ref{fgr:nitride} (a), the refractive index decreases with increasing \ce{N} content, in good agreement with previous work \cite{Gorin2008}. The refractive index determines the mode area for a guided mode in the $\alpha$\ce{SiNx} film. A higher refractive index allows for tighter modal confinement and overlap with a dipole (in this case an NV centre) located in the centre of the waveguide. This ensures that a large fraction of the emission is funnelled into single-mode waveguide. Since the refractive index of the film is lowered from $\sim 2$ to $\sim 1.9$ by increasing the N content, this will result in a slightly lower coupling efficiency. 

To measure the background PL spectrum of the different $\alpha$\ce{SiNx} films, we use a standard confocal microscope (see Supporting Information). In Fig \ref{fgr:nitride} (b), we plot the integrated PL spectra for different films as a function of gas ratio.  For reference, the integrated PL spectrum obtained from a bare fused silica substrate (green) and stoichiometric LPCVD \ce{Si3N4} films on Si (orange) are also shown. Two representative spectra, corresponding to low and high N content are shown in the figure inset. 
As Fig \ref{fgr:nitride} (b) demonstrates, the background PL emission can be reduced by almost two orders of magnitude, whereas the refractive index of the films drops by less than $5\%$, making it feasible for $\alpha$\ce{SiNx} to be used as a viable quantum photonics platform. 

\begin{figure}
 \includegraphics[width=0.45\textwidth]{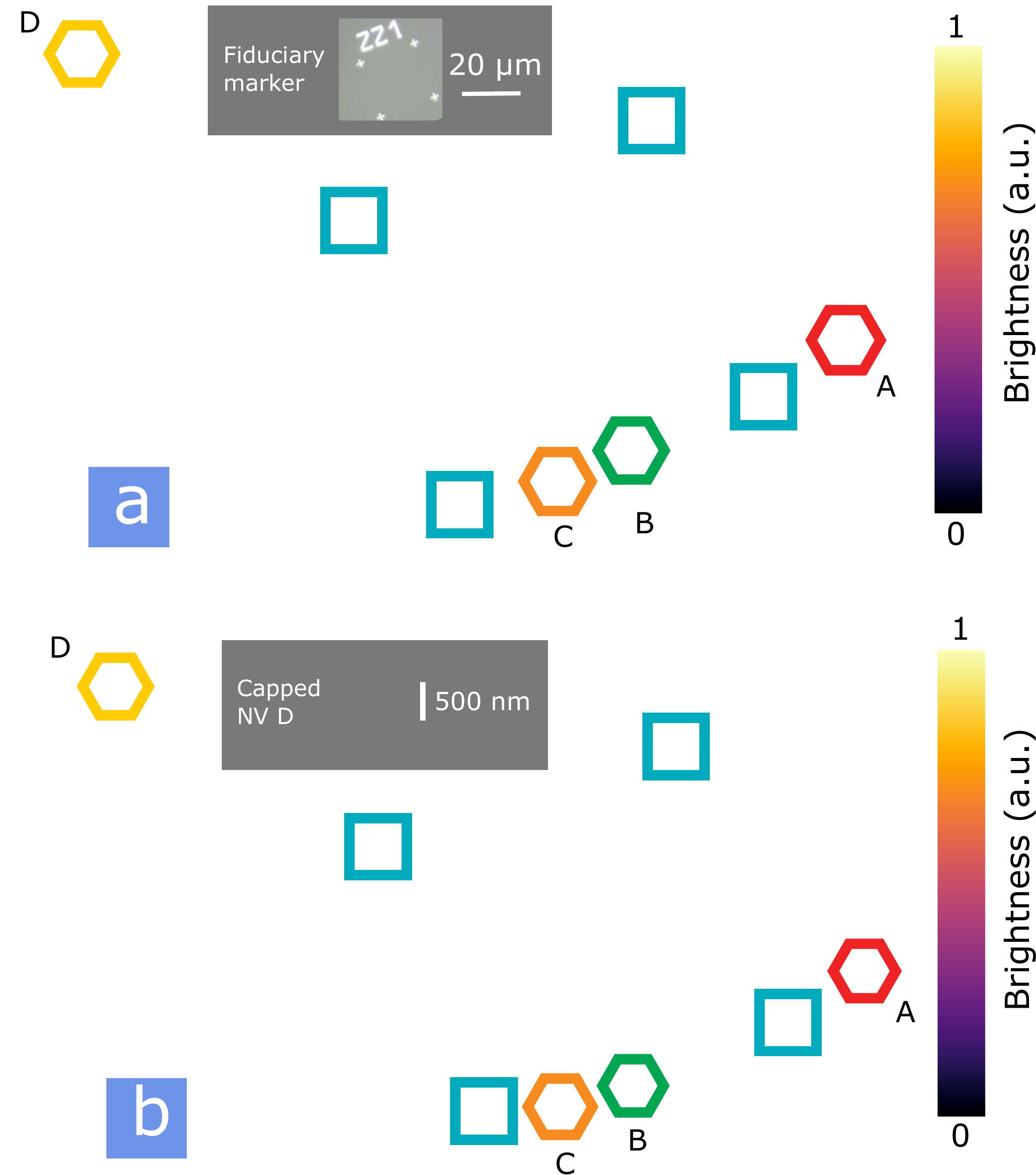}
  \caption{a) Large area confocal scan of a bare fused silica substrate coated with nanodiamonds containing NV centres. Pixel colour indicates the strength of the PL signal (cf. colourbar). The large bright spots, shown with square boxes, correspond to fiduciary markers (shown in the inset) to identify the positions of the NV centres. b) Confocal scan of the same area after encapsulation with 100 nm of low auto-fluorescence silicon nitride. The positions of the markers and the NV centres are again indicated. A high-resolution scan around one of the capped NV centres is shown in the inset.}
  \label{fgr:confocal}
\end{figure}


Rather than blindly probing the encapsulated film, we borrow techniques from localisation microscopy \cite{sapienza2015nanoscale,thompson2002precise} to record both the optical and spin properties of the same NV centre before and after nitride film encapsulation. Returning to a pre-characterised NV centre as opposed to measuring an ensemble of random NVs, we can better understand the effects of film encapsulation. Localisation also provides unambiguous proof that the NV centres isolated in nanodiamonds survive nanofabrication processes, in particular, exposure to plasma chemical vapour deposition as direct bombardment of NV centres with ions in sub-micron volumes is known to cause their destruction \cite{babinec2012topics}. A representative scan, on a bare fused silica substrate spin-coated with nanodiamonds, is shown in Fig \ref{fgr:confocal} (a).  To achieve localisation, we use fiduciary markers (crosses) fabricated using electron-beam lithography (indicated by rectangular boxes) with the spatial coordinates of the PL registered with respect to the alignment markers. High purity nanodiamonds are used with low nitrogen content to ensure a high probability of containing single NVs. To understand the effect of the $\alpha$\ce{SiNx} films on the optical and spin properties of the nanodiamonds, we pick four representative NV centres (A-D, labelled by hexagons in Fig \ref{fgr:confocal}(b)) with intensity auto-correlation $g^{(2)}(\tau)$ confirming single photon statistics. 

We encapsulate the sample with a low auto-fluorescence nitride film ($r =3$ in Fig \ref{fgr:nitride}). The confocal PL map of the same area for the encapsulated films is shown in Fig \ref{fgr:confocal} (b). For ease of comparison, the location of the alignment markers and the four pre-characterised single NV centres (A-D) are shown. It can be clearly seen, from the four bright spots, that the NV centre fluorescence is preserved after film encapsulation. The positions of the NV centres are also identical with respect to the markers pre- and post- film deposition, which confirms that the nanodiamonds have not been disturbed during the process of NV pre-characterisation, moving the sample from the lab to the cleanroom, film deposition and post-deposition characterisation back in the lab. A zoomed-in scan of the emission of NV D after nitride deposition is shown in the inset. The nitrogen rich silicon nitride films show observable bleaching effects. We do not fully understand the origin of this, but believe it is due to the presence of unpassivated charge traps in the amorphous nitride matrix, which are quenched by laser exposure.

\begin{figure}
   \includegraphics[width=0.5\textwidth]{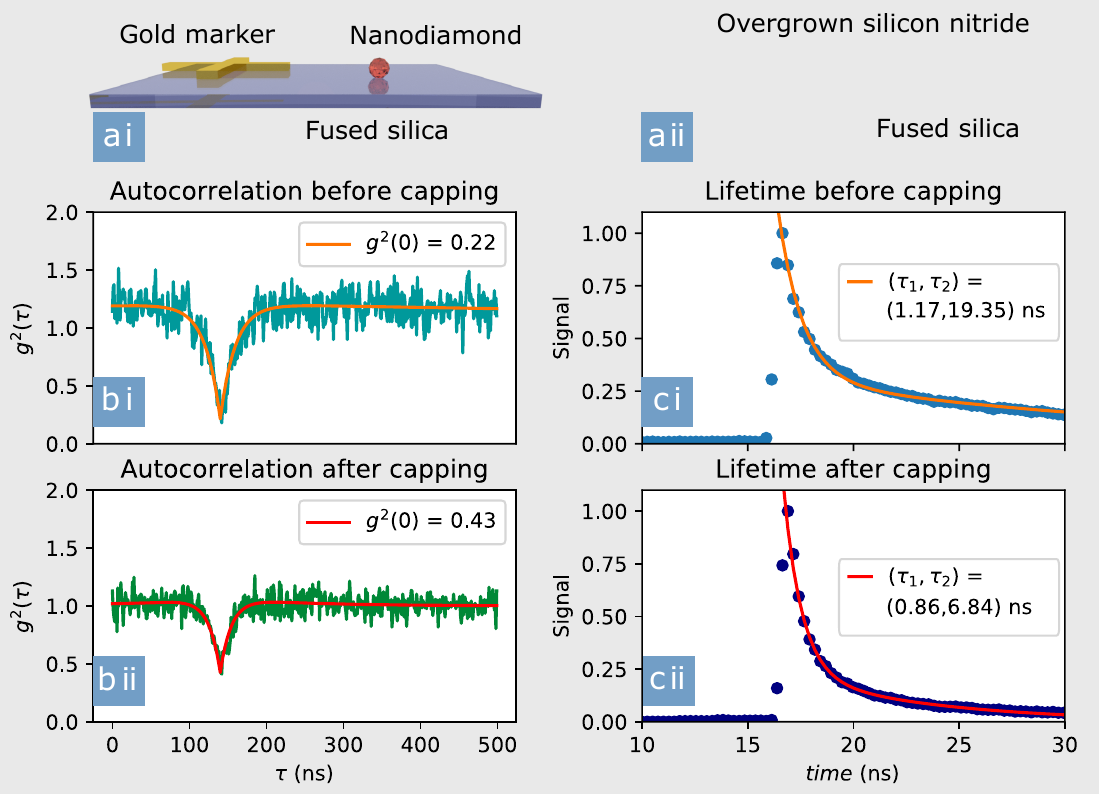}
  \caption{Optical characterisation of a representative single NV centre (NV D in Fig \ref{fgr:confocal}) measured i) before and ii) after encapsulation with a low auto-fluorescence nitride film. b) Intensity autocorrelation ($g^{(2)}(\tau)$) measurements showing a $g^{(2)}(0)$ < 0.5 preserved.  c) Fluorescence lifetime measurements showing that the fluorescence lifetime ($\tau_{2}$) is reduced by a factor of 3, attributed to radiative rate enhancement by coupling to slab modes in the encapsulating film.  }
  \label{fgr:photonstats}
\end{figure}

To quantify the effects of the nitride film on NV spin and optical coherence, we study the properties of isolated NV centres before and after film capping. We start by characterising the effect on optical properties. By measuring the intensity autocorrelation split between two detectors, $g^{(2)}(\tau) = \langle I(t)I(t+\tau)\rangle/\langle I(t)\rangle^2$, we can demonstrate that the emission is antibunched ($g^{(2)}(0) < 1$), corresponding to emission from a single quantum emitter. The photon statistics, displayed in Fig \ref{fgr:photonstats} (a) are not corrected for dark counts in the detectors. We fit an uncorrected $g^{(2)}(0) = 0.22$ for NV D measured on the bare fused silica substrate. Following encapsulation by a 100 nm low auto-fluorescence silicon nitride film ($r=3$ in Fig \ref{fgr:nitride}), we repeat the same measurement on NV D and find a $g^{(2)}(0) = 0.43$ observing anti-bunched emission from the encapsulated NV centre. In addition, $g^{(2)}(0) < 0.5$, which is the threshold for single emitter emission. The $g^{(2)}(0)$ has increased post-encapsulation which correlates with the observed background PL in Fig \ref{fgr:confocal}. Although the background fluorescence might be lower in other materials such as bulk diamond, a silicon nitride platform is clearly much closer to existing mature silicon technologies in terms of complexity, and here surpasses the threshold for quantum photonics. Antibunching is also observed for NV A-C following encapsulation, with uncorrected $g^{(2)}(0) = $0.52, 0.82 and 0.95 respectively (uncapped 0.17, 0.39, 0.09). NV A and D prove this platform is viable but we need further statistics to fully characterise the effect of encapsulation. To build a viable quantum photonics platform in silicon nitride, it is key that single emitters encapsulated in dielectric films can be identified and their emission statistics quantified. Our result provides a proof-of-principle demonstration of this idea. Amorphous dielectric films generally harbour traps and surface states which affect the energy levels of near-surface NV centres and could affect its fluorescence lifetime \cite{riedel2017deterministic}. In Fig \ref{fgr:photonstats}(b), the excited state fluorescence lifetime of the emitter located on a bare fused silica substrate is shown. The emission is fit by a bi-exponential function, $I(t) = A\tau_1\exp(-t/\tau_1) + B\tau_2\exp(-t/\tau_2)$, with a rapid $\tau_1 = 1.17$ ns decay attributed to the background, followed by the slower NV fluorescence, $\tau_2 = 19.35$ ns, characteristic of NVs in nanodiamonds \cite{beveratos2001nonclassical}. The contribution of the terms is proportional to the bare SNR observed in the confocal data (see Fig \ref{fgr:confocal}). After the NV centre is encapsulated, the contribution from the $\tau_1$ term increases, as expected from the decrease in SNR. The fluorescence lifetime $\tau_2$ decreased by a factor of 3 to $\tau_2 = 6.84$ ns. This is expected as the higher refractive index surrounding the NV centre increases the bare spontaneous emission rate by funnelling emission into slab waveguide modes in the film. To confirm this hypothesis, we need to map the dipole orientation of the NV centre and calculate the mode overlap with the guided modes supported by the film. We need to move to cryogenic temperatures below Jahn-Teller dominated dephasing to fully characterise the optical coherence of the NV centres for spin-photon applications \cite{fu2009observation}.

\begin{figure}
 \includegraphics[width=0.5\textwidth]{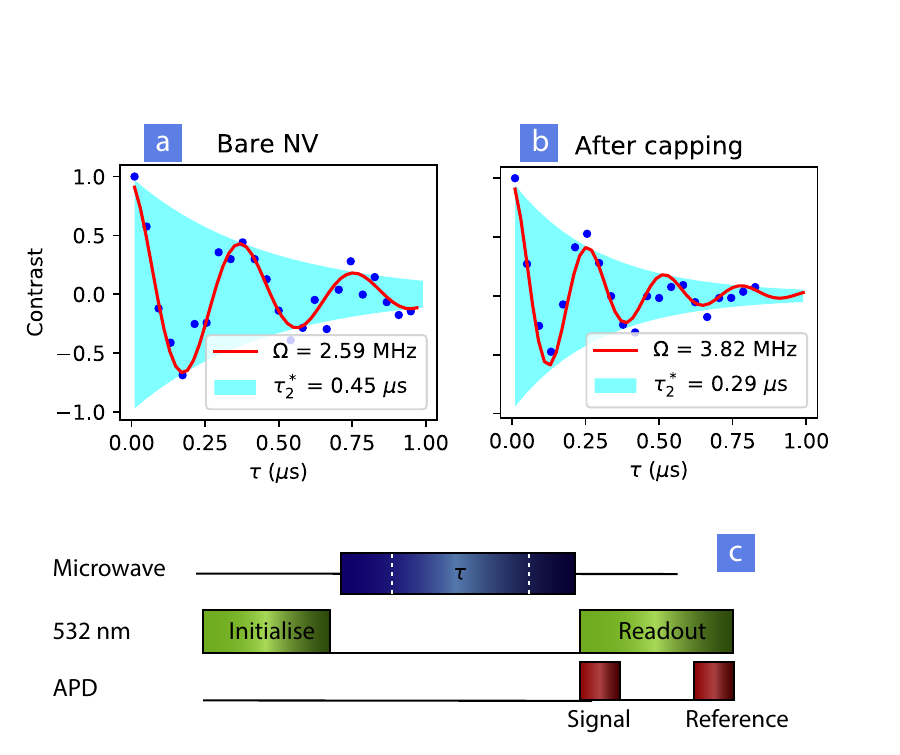}
  \caption{Free induction decay of NV D (from Fig \ref{fgr:confocal}) (a) before and (b) after encapsulation with the silicon nitride film. The fitted Rabi decay indicates the NV centre in nanodiamond maintains electron spin coherence in the nitride-rich silicon nitride environment. The diagram below indicates the laser and microwave pulse sequence used to carry out this spin coherence measurement.}
  \label{fgr:rabi}
\end{figure}

As discussed, what makes the NV centre attractive as a quantum information platform is its use as a spin-photon interface via cycling transitions and electron spin dependent fluorescence. Therefore, it becomes critical to quantify the effects of the film encapsulation on the electron spin associated with an individual NV centre.  To characterise the spin coherence, we perform a free induction decay measurement and extract the spin dephasing time $\tau_2^*$. The pulse sequence used to measure $\tau_2^*$ is shown in Fig \ref{fgr:rabi} (b). The NV centre located on the bare fused silica has a spin decay of $\tau_2^*$ = 0.45 ${ \mu}s$ , consistent with values reported for nanodiamonds in the literature with this fast measurement limited to noise from neighbouring free electrons \cite{knowles2014observing}. After capping with the silicon nitride, we can observe coherent Rabi oscillations and extract a spin decay time of $\tau_2^*$ = 0.29 ${ \mu}s$. Although $\tau_2^*$ has decreased, possibly due to unpassivated surface charges in the silicon nitride, it is still long lived enough to use dynamical decoupling pulses to refocus this spin and allow orders of magnitude higher coherences \cite{knowles2014observing}, a key technique for quantum memories implemented with NV centres . The Rabi frequency ($\Omega$) of the NV centre increases by a factor of 1.5 after encapsulation, this is likely due to repositioning the sample with respect to the RF antenna. We were unable to measure Rabi oscillations from NV A-C after encapsulation. On the other hand, NV D proves this platform is compatible with single spin measurements and we will move forward to collect statistics on large numbers of encapsulated NV centres, achievable by positioning nanodiamonds in arrays \cite{kianinia2016robust}. \\

We have conclusively demonstrated that nitrogen-rich silicon nitride serves as a viable quantum photonic platform for building scalable spin-photon interfaces around NV centres in nanodiamonds. Our experiments demonstrate that both the optical and spin properties of an individual NV centre is preserved after film encapsulation surpassing the single photon and single spin threshold. The encapsulated emitter offers a level of technological capability beyond monolithic NV centres in diamond, just as the solid state emitter offers to the trapped atom. Given the variety of high-performance photonic components already demonstrated in silicon nitride, our work opens up a promising route to build efficient visible photonic interfaces between solid state emitters. While we have focused on the NV centre in diamond, the work can be easily extended to other quantum photonic platforms under investigation in the visible regime, including perovskite quantum dots \cite{utzat2019coherent} and 2D transition metal dichalcogenides \cite{aharonovich2016solid}. 

See the Supporting Information for further experimental details.

\begin{acknowledgements}

The work was supported by the British Council IL6 project (352345416) and JGR's EPSRC fellowship (EP/M024458/1). Electron beam lithography and film deposition were carried out on equipment purchased through the Quantum Technology Capital QUPIC EPSRC grant (EP/N015126/1). JS and JMR are supported by the EPSRC Quantum Engineering Centre for Doctoral Training (EP/L015730/1). JMR acknowledges financial support from Consejo Nacional de Ciencia y Tecnologia (CONACyT). We would like to acknowledge  J. Barreto, M. Cryan, P. Jiang, D. McCutcheon, and A. Murray for valuable discussions and suggestions.

\end{acknowledgements}




\bibliography{single-photon}

\providecommand{\latin}[1]{#1}
\makeatletter
\providecommand{\doi}
  {\begingroup\let\do\@makeother\dospecials
  \catcode`\{=1 \catcode`\}=2 \doi@aux}
\providecommand{\doi@aux}[1]{\endgroup\texttt{#1}}
\makeatother
\providecommand*\mcitethebibliography{\thebibliography}
\csname @ifundefined\endcsname{endmcitethebibliography}
  {\let\endmcitethebibliography\endthebibliography}{}
\begin{mcitethebibliography}{38}
\providecommand*\natexlab[1]{#1}
\providecommand*\mciteSetBstSublistMode[1]{}
\providecommand*\mciteSetBstMaxWidthForm[2]{}
\providecommand*\mciteBstWouldAddEndPuncttrue
  {\def\EndOfBibitem{\unskip.}}
\providecommand*\mciteBstWouldAddEndPunctfalse
  {\let\EndOfBibitem\relax}
\providecommand*\mciteSetBstMidEndSepPunct[3]{}
\providecommand*\mciteSetBstSublistLabelBeginEnd[3]{}
\providecommand*\EndOfBibitem{}
\mciteSetBstSublistMode{f}
\mciteSetBstMaxWidthForm{subitem}{(\alph{mcitesubitemcount})}
\mciteSetBstSublistLabelBeginEnd
  {\mcitemaxwidthsubitemform\space}
  {\relax}
  {\relax}

\bibitem[Wang \latin{et~al.}(2018)Wang, Paesani, Ding, Santagati, Skrzypczyk,
  Salavrakos, Tura, Augusiak, Man{\v{c}}inska, Bacco, \latin{et~al.}
  others]{wang2018multidimensional}
Wang,~J.; Paesani,~S.; Ding,~Y.; Santagati,~R.; Skrzypczyk,~P.; Salavrakos,~A.;
  Tura,~J.; Augusiak,~R.; Man{\v{c}}inska,~L.; Bacco,~D., \latin{et~al.}
  Multidimensional quantum entanglement with large-scale integrated optics.
  \emph{Science} \textbf{2018}, \emph{360}, 285--291\relax
\mciteBstWouldAddEndPuncttrue
\mciteSetBstMidEndSepPunct{\mcitedefaultmidpunct}
{\mcitedefaultendpunct}{\mcitedefaultseppunct}\relax
\EndOfBibitem
\bibitem[Lindner and Rudolph(2009)Lindner, and Rudolph]{lindner2009proposal}
Lindner,~N.~H.; Rudolph,~T. Proposal for pulsed on-demand sources of photonic
  cluster state strings. \emph{Physical review letters} \textbf{2009},
  \emph{103}, 113602\relax
\mciteBstWouldAddEndPuncttrue
\mciteSetBstMidEndSepPunct{\mcitedefaultmidpunct}
{\mcitedefaultendpunct}{\mcitedefaultseppunct}\relax
\EndOfBibitem
\bibitem[Childress and Hanson(2013)Childress, and Hanson]{childress2013diamond}
Childress,~L.; Hanson,~R. Diamond NV centers for quantum computing and quantum
  networks. \emph{MRS bulletin} \textbf{2013}, \emph{38}, 134--138\relax
\mciteBstWouldAddEndPuncttrue
\mciteSetBstMidEndSepPunct{\mcitedefaultmidpunct}
{\mcitedefaultendpunct}{\mcitedefaultseppunct}\relax
\EndOfBibitem
\bibitem[Lodahl \latin{et~al.}(2015)Lodahl, Mahmoodian, and
  Stobbe]{lodahl2015interfacing}
Lodahl,~P.; Mahmoodian,~S.; Stobbe,~S. Interfacing single photons and single
  quantum dots with photonic nanostructures. \emph{Reviews of Modern Physics}
  \textbf{2015}, \emph{87}, 347\relax
\mciteBstWouldAddEndPuncttrue
\mciteSetBstMidEndSepPunct{\mcitedefaultmidpunct}
{\mcitedefaultendpunct}{\mcitedefaultseppunct}\relax
\EndOfBibitem
\bibitem[Awschalom \latin{et~al.}(2018)Awschalom, Hanson, Wrachtrup, and
  Zhou]{awschalom2018quantum}
Awschalom,~D.~D.; Hanson,~R.; Wrachtrup,~J.; Zhou,~B.~B. Quantum technologies
  with optically interfaced solid-state spins. \emph{Nature Photonics}
  \textbf{2018}, \emph{12}, 516\relax
\mciteBstWouldAddEndPuncttrue
\mciteSetBstMidEndSepPunct{\mcitedefaultmidpunct}
{\mcitedefaultendpunct}{\mcitedefaultseppunct}\relax
\EndOfBibitem
\bibitem[Xia \latin{et~al.}(2014)Xia, Wang, Xiao, Dubey, and
  Ramasubramaniam]{xia2014two}
Xia,~F.; Wang,~H.; Xiao,~D.; Dubey,~M.; Ramasubramaniam,~A. Two-dimensional
  material nanophotonics. \emph{Nature Photonics} \textbf{2014}, \emph{8},
  899\relax
\mciteBstWouldAddEndPuncttrue
\mciteSetBstMidEndSepPunct{\mcitedefaultmidpunct}
{\mcitedefaultendpunct}{\mcitedefaultseppunct}\relax
\EndOfBibitem
\bibitem[Polisseni \latin{et~al.}(2016)Polisseni, Major, Boissier, Grandi,
  Clark, and Hinds]{Polisseni:16}
Polisseni,~C.; Major,~K.~D.; Boissier,~S.; Grandi,~S.; Clark,~A.~S.;
  Hinds,~E.~A. Stable, single-photon emitter in a thin organic crystal for
  application to quantum-photonic devices. \emph{Opt. Express} \textbf{2016},
  \emph{24}, 5615--5627\relax
\mciteBstWouldAddEndPuncttrue
\mciteSetBstMidEndSepPunct{\mcitedefaultmidpunct}
{\mcitedefaultendpunct}{\mcitedefaultseppunct}\relax
\EndOfBibitem
\bibitem[Zhong \latin{et~al.}(2015)Zhong, Kindem, Miyazono, and
  Faraon]{zhong2015nanophotonic}
Zhong,~T.; Kindem,~J.~M.; Miyazono,~E.; Faraon,~A. Nanophotonic coherent
  light--matter interfaces based on rare-earth-doped crystals. \emph{Nature
  communications} \textbf{2015}, \emph{6}, 8206\relax
\mciteBstWouldAddEndPuncttrue
\mciteSetBstMidEndSepPunct{\mcitedefaultmidpunct}
{\mcitedefaultendpunct}{\mcitedefaultseppunct}\relax
\EndOfBibitem
\bibitem[Knowles \latin{et~al.}(2014)Knowles, Kara, and
  Atat{\"u}re]{knowles2014observing}
Knowles,~H.~S.; Kara,~D.~M.; Atat{\"u}re,~M. Observing bulk diamond spin
  coherence in high-purity nanodiamonds. \emph{Nature materials} \textbf{2014},
  \emph{13}, 21\relax
\mciteBstWouldAddEndPuncttrue
\mciteSetBstMidEndSepPunct{\mcitedefaultmidpunct}
{\mcitedefaultendpunct}{\mcitedefaultseppunct}\relax
\EndOfBibitem
\bibitem[Hirose and Cappellaro(2016)Hirose, and Cappellaro]{hirose2016coherent}
Hirose,~M.; Cappellaro,~P. Coherent feedback control of a single qubit in
  diamond. \emph{Nature} \textbf{2016}, \emph{532}, 77\relax
\mciteBstWouldAddEndPuncttrue
\mciteSetBstMidEndSepPunct{\mcitedefaultmidpunct}
{\mcitedefaultendpunct}{\mcitedefaultseppunct}\relax
\EndOfBibitem
\bibitem[Bernien \latin{et~al.}(2013)Bernien, Hensen, Pfaff, Koolstra, Blok,
  Robledo, Taminiau, Markham, Twitchen, Childress, and
  Hanson]{bernien2013heralded}
Bernien,~H.; Hensen,~B.; Pfaff,~W.; Koolstra,~G.; Blok,~M.; Robledo,~L.;
  Taminiau,~T.; Markham,~M.; Twitchen,~D.; Childress,~L.; Hanson,~R. Heralded
  entanglement between solid-state qubits separated by three metres.
  \emph{Nature} \textbf{2013}, \emph{497}, 86\relax
\mciteBstWouldAddEndPuncttrue
\mciteSetBstMidEndSepPunct{\mcitedefaultmidpunct}
{\mcitedefaultendpunct}{\mcitedefaultseppunct}\relax
\EndOfBibitem
\bibitem[Faraon \latin{et~al.}(2012)Faraon, Santori, Huang, Acosta, and
  Beausoleil]{faraon2012coupling}
Faraon,~A.; Santori,~C.; Huang,~Z.; Acosta,~V.~M.; Beausoleil,~R.~G. Coupling
  of nitrogen-vacancy centers to photonic crystal cavities in monocrystalline
  diamond. \emph{Physical review letters} \textbf{2012}, \emph{109},
  033604\relax
\mciteBstWouldAddEndPuncttrue
\mciteSetBstMidEndSepPunct{\mcitedefaultmidpunct}
{\mcitedefaultendpunct}{\mcitedefaultseppunct}\relax
\EndOfBibitem
\bibitem[Sipahigil \latin{et~al.}(2016)Sipahigil, Evans, Sukachev, Burek,
  Borregaard, Bhaskar, Nguyen, Pacheco, Atikian, Meuwly, \latin{et~al.}
  others]{sipahigil2016integrated}
Sipahigil,~A.; Evans,~R.~E.; Sukachev,~D.~D.; Burek,~M.~J.; Borregaard,~J.;
  Bhaskar,~M.~K.; Nguyen,~C.~T.; Pacheco,~J.~L.; Atikian,~H.~A.; Meuwly,~C.,
  \latin{et~al.}  An integrated diamond nanophotonics platform for
  quantum-optical networks. \emph{Science} \textbf{2016}, \emph{354},
  847--850\relax
\mciteBstWouldAddEndPuncttrue
\mciteSetBstMidEndSepPunct{\mcitedefaultmidpunct}
{\mcitedefaultendpunct}{\mcitedefaultseppunct}\relax
\EndOfBibitem
\bibitem[Choy \latin{et~al.}(2012)Choy, Bradley, Deotare, Burgess, Evans,
  Mazur, and Lon{\v{c}}ar]{choy2012integrated}
Choy,~J.~T.; Bradley,~J.~D.; Deotare,~P.~B.; Burgess,~I.~B.; Evans,~C.~C.;
  Mazur,~E.; Lon{\v{c}}ar,~M. Integrated TiO 2 resonators for visible
  photonics. \emph{Optics letters} \textbf{2012}, \emph{37}, 539--541\relax
\mciteBstWouldAddEndPuncttrue
\mciteSetBstMidEndSepPunct{\mcitedefaultmidpunct}
{\mcitedefaultendpunct}{\mcitedefaultseppunct}\relax
\EndOfBibitem
\bibitem[Pernice \latin{et~al.}(2012)Pernice, Xiong, and Tang]{pernice2012high}
Pernice,~W.~H.; Xiong,~C.; Tang,~H.~X. High Q micro-ring resonators fabricated
  from polycrystalline aluminum nitride films for near infrared and visible
  photonics. \emph{Optics express} \textbf{2012}, \emph{20}, 12261--12269\relax
\mciteBstWouldAddEndPuncttrue
\mciteSetBstMidEndSepPunct{\mcitedefaultmidpunct}
{\mcitedefaultendpunct}{\mcitedefaultseppunct}\relax
\EndOfBibitem
\bibitem[Kippenberg \latin{et~al.}(2011)Kippenberg, Holzwarth, and
  Diddams]{kippenberg2011microresonator}
Kippenberg,~T.~J.; Holzwarth,~R.; Diddams,~S.~A. Microresonator-based optical
  frequency combs. \emph{science} \textbf{2011}, \emph{332}, 555--559\relax
\mciteBstWouldAddEndPuncttrue
\mciteSetBstMidEndSepPunct{\mcitedefaultmidpunct}
{\mcitedefaultendpunct}{\mcitedefaultseppunct}\relax
\EndOfBibitem
\bibitem[Li \latin{et~al.}(2016)Li, Davan{\c{c}}o, and
  Srinivasan]{li2016efficient}
Li,~Q.; Davan{\c{c}}o,~M.; Srinivasan,~K. Efficient and low-noise
  single-photon-level frequency conversion interfaces using silicon
  nanophotonics. \emph{Nature Photonics} \textbf{2016}, \emph{10}, 406\relax
\mciteBstWouldAddEndPuncttrue
\mciteSetBstMidEndSepPunct{\mcitedefaultmidpunct}
{\mcitedefaultendpunct}{\mcitedefaultseppunct}\relax
\EndOfBibitem
\bibitem[Domenech \latin{et~al.}(2018)Domenech, Porcel, Jans, Hoofman,
  Geuzebroek, Dumon, van~der Vliet, Witzens, Bourguignon, Artundo,
  \latin{et~al.} others]{domenech2018pix4life}
Domenech,~J.~D.; Porcel,~M.~A.; Jans,~H.; Hoofman,~R.; Geuzebroek,~D.;
  Dumon,~P.; van~der Vliet,~M.; Witzens,~J.; Bourguignon,~E.; Artundo,~I.,
  \latin{et~al.}  PIX4life: photonic integrated circuits for bio-photonics.
  Integrated Photonics Research, Silicon and Nanophotonics. 2018; pp
  ITh3B--1\relax
\mciteBstWouldAddEndPuncttrue
\mciteSetBstMidEndSepPunct{\mcitedefaultmidpunct}
{\mcitedefaultendpunct}{\mcitedefaultseppunct}\relax
\EndOfBibitem
\bibitem[Stroganov and Geiselmann(2019)Stroganov, and
  Geiselmann]{stroganovsilicon}
Stroganov,~A.; Geiselmann,~M. Silicon Nitride PICs Platform Development from a
  Foundry Perspective: From Concepts to Real Applications. 2019\relax
\mciteBstWouldAddEndPuncttrue
\mciteSetBstMidEndSepPunct{\mcitedefaultmidpunct}
{\mcitedefaultendpunct}{\mcitedefaultseppunct}\relax
\EndOfBibitem
\bibitem[Mouradian \latin{et~al.}(2015)Mouradian, Schr{\"o}der, Poitras, Li,
  Goldstein, Chen, Walsh, Cardenas, Markham, Twitchen, Lipson, and
  Englund]{mouradian2015scalable}
Mouradian,~S.~L.; Schr{\"o}der,~T.; Poitras,~C.~B.; Li,~L.; Goldstein,~J.;
  Chen,~E.~H.; Walsh,~M.; Cardenas,~J.; Markham,~M.~L.; Twitchen,~D.;
  Lipson,~M.; Englund,~D. Scalable integration of long-lived quantum memories
  into a photonic circuit. \emph{Physical Review X} \textbf{2015}, \emph{5},
  031009\relax
\mciteBstWouldAddEndPuncttrue
\mciteSetBstMidEndSepPunct{\mcitedefaultmidpunct}
{\mcitedefaultendpunct}{\mcitedefaultseppunct}\relax
\EndOfBibitem
\bibitem[Cernansky \latin{et~al.}(2018)Cernansky, Martini, and
  Politi]{Cernansky2018}
Cernansky,~R.; Martini,~F.; Politi,~A. {Complementary metal-oxide semiconductor
  compatible source of single photons at near-visible wavelengths}.
  \emph{Optics Letters} \textbf{2018}, \emph{43}, 855\relax
\mciteBstWouldAddEndPuncttrue
\mciteSetBstMidEndSepPunct{\mcitedefaultmidpunct}
{\mcitedefaultendpunct}{\mcitedefaultseppunct}\relax
\EndOfBibitem
\bibitem[Gorin \latin{et~al.}(2008)Gorin, Jaouad, Grondin, Aimez, and
  Charette]{Gorin2008}
Gorin,~A.; Jaouad,~A.; Grondin,~E.; Aimez,~V.; Charette,~P. {Fabrication of
  silicon nitride waveguides for visible-light using PECVD: a study of the
  effect of plasma frequency on optical properties}. \emph{Optics Express}
  \textbf{2008}, \emph{16}, 13509\relax
\mciteBstWouldAddEndPuncttrue
\mciteSetBstMidEndSepPunct{\mcitedefaultmidpunct}
{\mcitedefaultendpunct}{\mcitedefaultseppunct}\relax
\EndOfBibitem
\bibitem[Schell \latin{et~al.}(2013)Schell, Kaschke, Fischer, Henze, Wolters,
  Wegener, and Benson]{schell2013three}
Schell,~A.~W.; Kaschke,~J.; Fischer,~J.; Henze,~R.; Wolters,~J.; Wegener,~M.;
  Benson,~O. Three-dimensional quantum photonic elements based on single
  nitrogen vacancy-centres in laser-written microstructures. \emph{Scientific
  reports} \textbf{2013}, \emph{3}, 1577\relax
\mciteBstWouldAddEndPuncttrue
\mciteSetBstMidEndSepPunct{\mcitedefaultmidpunct}
{\mcitedefaultendpunct}{\mcitedefaultseppunct}\relax
\EndOfBibitem
\bibitem[Khalid \latin{et~al.}(2015)Khalid, Chung, Rajasekharan, Lau, Karle,
  Gibson, and Tomljenovic-Hanic]{khalid2015lifetime}
Khalid,~A.; Chung,~K.; Rajasekharan,~R.; Lau,~D.~W.; Karle,~T.~J.;
  Gibson,~B.~C.; Tomljenovic-Hanic,~S. Lifetime reduction and enhanced emission
  of single photon color centers in nanodiamond via surrounding refractive
  index modification. \emph{Scientific reports} \textbf{2015}, \emph{5},
  11179\relax
\mciteBstWouldAddEndPuncttrue
\mciteSetBstMidEndSepPunct{\mcitedefaultmidpunct}
{\mcitedefaultendpunct}{\mcitedefaultseppunct}\relax
\EndOfBibitem
\bibitem[Hui \latin{et~al.}(2019)Hui, Chen, Azuma, Chang, Hsieh, and
  Chang]{hui2019all}
Hui,~Y.~Y.; Chen,~O.~Y.; Azuma,~T.; Chang,~B.-M.; Hsieh,~F.-J.; Chang,~H.-C.
  All-Optical Thermometry with Nitrogen-Vacancy Centers in Nanodiamond-Embedded
  Polymer Films. \emph{The Journal of Physical Chemistry C} \textbf{2019},
  \relax
\mciteBstWouldAddEndPunctfalse
\mciteSetBstMidEndSepPunct{\mcitedefaultmidpunct}
{}{\mcitedefaultseppunct}\relax
\EndOfBibitem
\bibitem[Van~de Ven \latin{et~al.}(1990)Van~de Ven, Connick, and
  Harrus]{van1990advantages}
Van~de Ven,~E.~P.; Connick,~I.-W.; Harrus,~A.~S. Advantages of dual frequency
  PECVD for deposition of ILD and passivation films. Seventh International IEEE
  Conference on VLSI Multilevel Interconnection. 1990; pp 194--201\relax
\mciteBstWouldAddEndPuncttrue
\mciteSetBstMidEndSepPunct{\mcitedefaultmidpunct}
{\mcitedefaultendpunct}{\mcitedefaultseppunct}\relax
\EndOfBibitem
\bibitem[Sapienza \latin{et~al.}(2015)Sapienza, Davan{\c{c}}o, Badolato, and
  Srinivasan]{sapienza2015nanoscale}
Sapienza,~L.; Davan{\c{c}}o,~M.; Badolato,~A.; Srinivasan,~K. Nanoscale optical
  positioning of single quantum dots for bright and pure single-photon
  emission. \emph{Nature communications} \textbf{2015}, \emph{6}, 7833\relax
\mciteBstWouldAddEndPuncttrue
\mciteSetBstMidEndSepPunct{\mcitedefaultmidpunct}
{\mcitedefaultendpunct}{\mcitedefaultseppunct}\relax
\EndOfBibitem
\bibitem[Thompson \latin{et~al.}(2002)Thompson, Larson, and
  Webb]{thompson2002precise}
Thompson,~R.~E.; Larson,~D.~R.; Webb,~W.~W. Precise nanometer localization
  analysis for individual fluorescent probes. \emph{Biophysical journal}
  \textbf{2002}, \emph{82}, 2775--2783\relax
\mciteBstWouldAddEndPuncttrue
\mciteSetBstMidEndSepPunct{\mcitedefaultmidpunct}
{\mcitedefaultendpunct}{\mcitedefaultseppunct}\relax
\EndOfBibitem
\bibitem[Babinec(2012)]{babinec2012topics}
Babinec,~T.~M. Topics in Nanophotonic Devices for Nitrogen-Vacancy Color
  Centers in Diamond. \emph{ProQuest LLC} \textbf{2012}, \relax
\mciteBstWouldAddEndPunctfalse
\mciteSetBstMidEndSepPunct{\mcitedefaultmidpunct}
{}{\mcitedefaultseppunct}\relax
\EndOfBibitem
\bibitem[Riedel \latin{et~al.}(2017)Riedel, S{\"o}llner, Shields, Starosielec,
  Appel, Neu, Maletinsky, and Warburton]{riedel2017deterministic}
Riedel,~D.; S{\"o}llner,~I.; Shields,~B.~J.; Starosielec,~S.; Appel,~P.;
  Neu,~E.; Maletinsky,~P.; Warburton,~R.~J. Deterministic enhancement of
  coherent photon generation from a nitrogen-vacancy center in ultrapure
  diamond. \emph{Physical Review X} \textbf{2017}, \emph{7}, 031040\relax
\mciteBstWouldAddEndPuncttrue
\mciteSetBstMidEndSepPunct{\mcitedefaultmidpunct}
{\mcitedefaultendpunct}{\mcitedefaultseppunct}\relax
\EndOfBibitem
\bibitem[Beveratos \latin{et~al.}(2001)Beveratos, Brouri, Gacoin, Poizat, and
  Grangier]{beveratos2001nonclassical}
Beveratos,~A.; Brouri,~R.; Gacoin,~T.; Poizat,~J.-P.; Grangier,~P. Nonclassical
  radiation from diamond nanocrystals. \emph{Physical Review A} \textbf{2001},
  \emph{64}, 061802\relax
\mciteBstWouldAddEndPuncttrue
\mciteSetBstMidEndSepPunct{\mcitedefaultmidpunct}
{\mcitedefaultendpunct}{\mcitedefaultseppunct}\relax
\EndOfBibitem
\bibitem[Fu \latin{et~al.}(2009)Fu, Santori, Barclay, Rogers, Manson, and
  Beausoleil]{fu2009observation}
Fu,~K.-M.~C.; Santori,~C.; Barclay,~P.~E.; Rogers,~L.~J.; Manson,~N.~B.;
  Beausoleil,~R.~G. Observation of the dynamic Jahn-Teller effect in the
  excited states of nitrogen-vacancy centers in diamond. \emph{Physical Review
  Letters} \textbf{2009}, \emph{103}, 256404\relax
\mciteBstWouldAddEndPuncttrue
\mciteSetBstMidEndSepPunct{\mcitedefaultmidpunct}
{\mcitedefaultendpunct}{\mcitedefaultseppunct}\relax
\EndOfBibitem
\bibitem[Jelezko \latin{et~al.}(2004)Jelezko, Gaebel, Popa, Gruber, and
  Wrachtrup]{jelezko2004observation}
Jelezko,~F.; Gaebel,~T.; Popa,~I.; Gruber,~A.; Wrachtrup,~J. Observation of
  coherent oscillations in a single electron spin. \emph{Physical review
  letters} \textbf{2004}, \emph{92}, 076401\relax
\mciteBstWouldAddEndPuncttrue
\mciteSetBstMidEndSepPunct{\mcitedefaultmidpunct}
{\mcitedefaultendpunct}{\mcitedefaultseppunct}\relax
\EndOfBibitem
\bibitem[Kianinia \latin{et~al.}(2016)Kianinia, Shimoni, Bendavid, Schell,
  Randolph, Toth, Aharonovich, and Lobo]{kianinia2016robust}
Kianinia,~M.; Shimoni,~O.; Bendavid,~A.; Schell,~A.~W.; Randolph,~S.~J.;
  Toth,~M.; Aharonovich,~I.; Lobo,~C.~J. Robust, directed assembly of
  fluorescent nanodiamonds. \emph{Nanoscale} \textbf{2016}, \emph{8},
  18032--18037\relax
\mciteBstWouldAddEndPuncttrue
\mciteSetBstMidEndSepPunct{\mcitedefaultmidpunct}
{\mcitedefaultendpunct}{\mcitedefaultseppunct}\relax
\EndOfBibitem
\bibitem[Utzat \latin{et~al.}(2019)Utzat, Sun, Kaplan, Krieg, Ginterseder,
  Spokoyny, Klein, Shulenberger, Perkinson, Kovalenko, \latin{et~al.}
  others]{utzat2019coherent}
Utzat,~H.; Sun,~W.; Kaplan,~A.~E.; Krieg,~F.; Ginterseder,~M.; Spokoyny,~B.;
  Klein,~N.~D.; Shulenberger,~K.~E.; Perkinson,~C.~F.; Kovalenko,~M.~V.,
  \latin{et~al.}  Coherent single-photon emission from colloidal lead halide
  perovskite quantum dots. \emph{Science} \textbf{2019}, \emph{363},
  1068--1072\relax
\mciteBstWouldAddEndPuncttrue
\mciteSetBstMidEndSepPunct{\mcitedefaultmidpunct}
{\mcitedefaultendpunct}{\mcitedefaultseppunct}\relax
\EndOfBibitem
\bibitem[Aharonovich \latin{et~al.}(2016)Aharonovich, Englund, and
  Toth]{aharonovich2016solid}
Aharonovich,~I.; Englund,~D.; Toth,~M. Solid-state single-photon emitters.
  \emph{Nature Photonics} \textbf{2016}, \emph{10}, 631\relax
\mciteBstWouldAddEndPuncttrue
\mciteSetBstMidEndSepPunct{\mcitedefaultmidpunct}
{\mcitedefaultendpunct}{\mcitedefaultseppunct}\relax
\EndOfBibitem
\bibitem[Binder \latin{et~al.}(2017)Binder, Stark, Tomek, Scheuer, Frank,
  Jahnke, M{\"u}ller, Schmitt, Metsch, Unden, \latin{et~al.}
  others]{binder2017qudi}
Binder,~J.~M.; Stark,~A.; Tomek,~N.; Scheuer,~J.; Frank,~F.; Jahnke,~K.~D.;
  M{\"u}ller,~C.; Schmitt,~S.; Metsch,~M.~H.; Unden,~T., \latin{et~al.}  Qudi:
  A modular python suite for experiment control and data processing.
  \emph{SoftwareX} \textbf{2017}, \emph{6}, 85--90\relax
\mciteBstWouldAddEndPuncttrue
\mciteSetBstMidEndSepPunct{\mcitedefaultmidpunct}
{\mcitedefaultendpunct}{\mcitedefaultseppunct}\relax
\EndOfBibitem
\end{mcitethebibliography}

\end{document}